\documentclass[y4paper, 12pt]{article}

\begin{document}

\baselineskip 8.0mm
\thispagestyle{empty}

\centerline{\large\bf Estimations of changes of the Sun's mass and }
\centerline{\large\bf the gravitation constant from the modern observations }
\centerline{\large\bf of planets  and spacecraft}
\bigskip
\centerline{\bf E. V. Pitjeva$^{1*}$,  N. P. Pitjev$^{2**}$ }
\centerline{\it $^1$Institute of Applied Astronomy of Russian Academy of
Sciences, St.Petersburg}
\centerline{\it $^2$St.Petersburg State University, St.Petersburg}

\bigskip\noindent

More than 635 000 positional observations of planets and spacecraft of
different types, mostly radiotechnical ones (1961-2010), have been used
for estimating possible changes of the gravitation constant, the solar mass,
and semi-major axes of planets, as well as the value of the astronomical unit,
related to them. The analysis of the observations has been performed on the
basis of the EPM2010 ephemerides of IAA RAS in post-newtonian approximation.
The EPM ephemerides are computed by numerical integration of the equations of
motion of the nine major planets, the Sun, the Moon, asteroids and
trans-neptunian objects.

The estimation of change for the geliocentric gravitation constant $GM_{\odot}$
has been obtained
\centerline{$(\dot {GM_{\odot}}) /GM_{\odot} =
(-5.0 \pm 4.1)\cdot10^{-14} \quad \hbox{in year} \quad \ (3\sigma$).}

\noindent The positive century changes of semi-major axes $ \dot a_i/a_i$ have
been determined simultaneously for the planets Mercury, Venus, Mars, Jupiter,
Saturn provided with high-accuracy sets of the observations, as expected if the
geliocentric gravitation constant is decreasing in century wise.
Perhaps, loss of the mass of the Sun $M_{\odot}$ the produces change of
$GM_{\odot}$ due to the solar radiation and the solar wind compensated
partially by the matter dropping on the Sun.

It was been found from  the obtained $GM_{\odot}$ change and taking into
account the maximal limits of the possible $M_{\odot}$ change that the
gravitation constant $\dot G/G$ falls within the interval

\centerline{$ -4.2\cdot10^{-14}  < \dot G/G < +7.5\cdot10^{-14}$\quad \hbox{in year}}

\noindent with the 95$\%$ probability.

The astronomical unit (au) is only connected with the geliocentric gravitation
constant by its definition. The decrease of $GM_{\odot}$ obtained in this
paper should correspond to the century decrease of au. However, it has been
shown that the present accuracy level of observations does not permit to
evaluate the au change.
The attained possibility of fining the $GM_{\odot}$ change from high-accuracy
observations points that fixing the connection between $GM_{\odot}$ and au
at the certain time moment is desirable, as it is inconvenient highly
to have the changing value of the astronomical unit.



$^*$E-mail: {\tt evp@ipa.nw.ru}

$^{**}$E-mail: {\tt ai@astro.spbu.ru}
\vskip 3em \noindent

\centerline{\bf INTRODUCTION}
\smallskip

     The whole array of high-precision observations of the planets and
the development of modern planetary ephemeris form prerequisites for the study
of very subtle effects, in particular, a change of the geliocentric
gravitational constant $ GM_{\odot} $ with time. The question of variability 
and the possible rate of the change of gravitational constant $ G $ is 
regularly raised and considered in some cosmological theories (Uzan, 2003; 
2009). The Sun's mass $ M_{\odot} $, can not be absolutely constant too. On the 
one hand, it decreases due to continuous thermonuclear reactions and production 
of the radiant energy, with the matter carried away by the solar wind. On the 
other hand, there is a regular drop of interplanetary substances on the Sun, 
including dust, meteoroids, asteroids and comets.

     In the history of the Sun there have possibly been the periods of positive 
and negative changes of the solar mass. In the initial period, shortly after the 
formation of the protosun and the beginning of nuclear reactions in it,
the mass of the central body probably increased due to the continuing 
compression to the center of the initial cloud. During the formation of the
planetary system, until the interplanetary space was cleared, the mass of the
matter falling  on the Sun was higher than the mass of the Sun reduced due to
the light and corpuscular radiation. Now traces
of the original, protosolar cloud are only likely to remain on the periphery of 
the Solar system beyond Neptune, the issue of the quantifying
changes of the mass of the Sun still remains open because of the uncertainty
of the overall balance including the mass loss due to radiation and the matter
carried away by the solar wind, and the mass increase due to the matter falling 
on the Sun, in particular, comets whose collision with the photosphere of the 
Sun was repeatedly recorded by the SOHO space observatory 
(http://ares.nrl.navy.mil/sungrazer/).
Any estimation is difficult because of the complexity of reliable estimation
of falling matter mass as well as the time-varied intensity and the
angular distribution of the solar wind in space.
In this paper, we attempt to obtain experimental estimates of the change of the
solar mass, namely, the geliocentric gravitational constant $ GM_{\odot} $ from
the analysis of observational data of motion of planets and spacecraft.

    The change of the value of the astronomical unit (au) is related to the 
change of the geliocentric gravitational constant. The astronomical unit is
close in its magnitude to the average distance from the Earth to the Sun, but,
by its definition it is only related to the heliocentric gravitational constant
$ GM_{\odot} $. From the obtained estimate $\dot {GM_{\odot}} $, we can 
estimate the possible change in time of the au value. This estimate can be 
compared with the direct change value of au obtained from observations.

\bigskip
\centerline{\bf EXPECTED EFFECTS OF THE CHANGE OF THE SUN'S MASS}
\smallskip

     Estimations of the mass change of the Sun and its rate repeatedly are 
cited in the papers related to the exploration of solar physics, the solar wind 
and radiation (e.g. Sunyaev, 1986; Livingston, 2000). The Sun's luminosity
$L_{\odot}$ somewhat varies  during the eleven year cycle and the $\sim$ 27-day
rotation around its axis; however, the fluctuations $L_{\odot} $ do not exceed
$0.1 \div 0.2\% $ (Frohlich, Lean, 1998; 2004). If we take the average total
solar luminosity to be $L_{\odot} = 3.846 \cdot10^{33}$ erg/s and the mass of 
the Sun $M_{\odot} = 1.9891 \cdot10^{33}$ g (Brun et al, 1998), then the 
decrease of the mass of the Sun due to radiation as a fraction of the solar mass 
is equal to $\dot M_{\odot} = -6.789 \cdot10^{-14} M_{\odot}$ per year.

     The mass carried away with the solar wind was also repeatedly evaluated.
The basic composition of the solar wind is as follows: approximately 95\% --- 
protons, 4\% --- nuclei of the helium atoms, and less than 1\% --- nuclei of 
atoms of other elements (C, N, O, Ne, Mg, Ca, Si, Fe) (Brandt, 1973; Hundhauzen, 
1976). The total number of particles flying away every second, is approximate by 
estimated as $1.3\cdot10^{36}$ (Kallenrode, 2004). The flow of the solar wind 
affects the activity of the Sun, coronal mass ejections. Typically, the average 
loss per year through the solar wind is estimated as $2\cdot10^{-14} M_{\odot}$ 
(Hundhauzen, 1976; Hundhausen, 1997; Meyer-Vernet, 2007), that is, less than a 
third of the mass loss due to radiation. There are estimates 
$ (2 \div 3)\cdot10^{-14} M_{\odot}$ per year (Sunyaev, 1986; Carroll, Ostlie, 
1996; Livingston, 2000), where a value of $ 3\cdot10^{-14} M_{\odot}$ can be 
considered the upper limit of the mass carried away by the solar wind. The 
cumulative effect of the relative annual decrease of the mass of the Sun due to 
radiation and the solar wind can be restricted by the inequality
\begin{equation}\label{f-1}
-9.8\cdot10^{-14} < \dot M_{\odot} / M_{\odot} < -8.8\cdot10^{-14}.
\end{equation}

     The reverse process occurs due to the fall of the dust, meteor, asteroid
and comet substance on the Sun. The dusty environment can not make a significant 
contribution to the mass of the material fallen. According to the current data, 
the density of the interplanetary dust decreases with distance from the Sun, so 
that at the distance greater than 3 au there is not dust practically, with
the 2/3ds of the interplanetary dust concentrating in particles of
$10^{-5} \div 10^{-3}$ g, and the size of dust particles being mostly
$1 \div 10 \ {\mu}m $ (Mann et al, 2010). The total mass of the dust matter is
estimated approximately as $10^{19} \div 10^{20}$ g (Sunyaev, 1986). Even with 
the assumption that all the mass reaches the Sun in several thousands years, the 
rate of the dust fall-out particles will be less than $10^{-16}M_{\odot}$ per 
year. However, the dust particles smaller than 2 microns are swept by the solar 
pressure, while the ones greater than 2 microns move towards the Sun. The most 
part of the approaching dust sublimates within 0.1 au ($\sim20 R_{\odot}$) and 
can not reach the surface of the Sun. It is also necessary to note, that a 
substantial part of the dust is carried away by the solar wind to the periphery 
of the Solar system (Mann et al., 2010). Therefore, the possible rate of the 
dust component falling out on the Sun is much smaller than 
$(10^{-16} \div 10^{-17})\cdot M_{\odot}$ per year.

    The larger particles, meteoroids and asteroids may fall on the Sun. Studies
show that there is a constant migration of asteroids with an opportunity to 
complete the orbit evolution by the collision with the Sun (Farinella and et al, 
1994; Gladman et al, 1997). The total number of small bodies is very large; the 
number of bodies larger than 1 ~km is about 1 million.  A significant part of 
asteroids move within a region close to the orbit planes of major planets of the 
Solar system, mostly situated in the belt between the orbits of Mars and 
Jupiter. The current estimates of the total mass of the asteroid belt give 
$(13 \pm 2)\cdot10^{-10} M_{\odot}$ (Pitjeva, 2010b), i.e. less than $10^{-3}$ 
mass of the Earth. For the ring to be able to exist for tens or hundreds 
millions of years, the fraction of the outgoing annually material should be 
significantly smaller than $10^{-7} \div 10^{-8}$ of the total mass of the 
asteroid belt, in case the main asteroid belt is not replenished from the 
outside. With the outgoing material falling on the Sun not regularly, we find 
that the upper limit of possible mass of the Sun drop-down material from the 
main belt is less than $(10^{-16} \div 10^{-17})\cdot M_{\odot}$ per year. Thus, 
we obtain a significantly lower value (two to three orders of magnitude) than 
the decrease of the solar mass by radiation and solar wind, so in the solar 
neighbourhood and in the field of the main asteroid belt there is no sufficient 
interplanetary matter migrating to the Sun to be compared with the decrease of
the solar mass due to radiation.

   The mass of the matter that can come from distant regions of the solar 
system, mainly in the form of comets (Bailey et al 1992), is more uncertain.
There are trans-Neptunian areas --- the Kuiper Belt, a cloud of the Hills, the 
Oort cloud. Currently, a large number of comets is detected in the immediate 
vicinity of the Sun (sungrazing comets) using the LASCO coronagraph  
(http://lasco-www.nrl.navy.mil/) installed at the SOHO solar space observatory 
(Marsden, 1989; 2005). Comets close to and often passing near the Sun are not 
long-living. They can disintegrate into fragments, or completely "fall apart". 
An example is a large family of Kreutz comets (Sekanina, Chodas, 2007). Some of 
them enter directly to the dense layers of the Sun. On the pictures of the SOHO 
observatory (http://sungrazer.nrl.navy.mil/index.php) the death of small 
fragments of comets in the solar photosphere is regularly recorded (kamikaze 
comet). The falling rocky bodies are more difficult to register, as
during their approach to the Sun the glowing gas tail like that of icy objects 
and comets, does not form and the rocky bodies remain invisible. The overall 
contribution of the visible and invisible objects may be significant, although 
a reliable estimate of the total mass of substance reaching the Sun is extremely 
difficult. Nevertheless, the upper limits to the total mass can be specified.
Statistics of comets discovered by the SOHO observatory gives about 500 comets
for $30 \div 35$ months, on the average $170 \div 200$ comets per year. Many of 
them completely evaporate during their passage through the lower layers of the 
solar corona. There are registered repeatedly instances when the comets reached 
the photosphere. We assume that all the detected comets ``vanished'' and their
mass increased the mass of the Sun, an overestimated value will result. The 
relatively small comets with the size from tens to hundreds meters are usually
recorded, but for the upper estimate, we assume that the diameters of their 
nuclei ($d_{com}$) are several kilometers (as for the comets, which approach the 
spacecraft). If we take the average value of $d_{com}=5$ km, the density of 
3 g/cm$^3$ and double the result due to the missing and invisible falling 
objects, the annual upper bound is
\begin{equation}\label{f-2}
\dot M_{\odot,com} / M_{\odot} < +3.2\cdot10^{-14}.
\end{equation}
     This estimate is comparable to values obtained for the mass loss of the
Sun due to radiation and the solar wind, although it seems to be overestimated, 
as the large falling objects have not been actually registered. This value can be 
considered the upper limit of the possible increase of the solar mass due to the 
material falling in the form of comets, meteors, asteroids and dust.

     Now one can point to the two sides of the common interval, probably 
overestimated, which should contain the value of $\dot M_{\odot} / M_{\odot}$. 
To obtain the lower limit, let us take the maximum loss estimate  due to the 
solar wind, and at the same time,  the zero drop of the material on the Sun. To 
find the upper limit, we use the maximum estimate (\ref{f-2}) for the material 
falling on the Sun and the assumption that there is no mass loss due to the 
solar wind. Then, we obtain
\begin{equation}\label{f-3}
 - 9.8\cdot10^{-14} < \dot M_{\odot} / M_{\odot} < -3.6\cdot10^{-14}\quad \hbox{¢ £®¤}.
\end{equation}
There are the restrictions to be kept in mind when trying to estimate the 
experimental changes of the Sun's mass.

\bigskip
{\bf THE INFLUENCE OF THE CHANGE OF THE SOLAR MASS ON ORBITAL ELEMENTS OF PLANETS}
\smallskip

   The change of the $ M_{\odot} $ solar mass must lead to the appearance of 
variations of elements of the planet orbits, but a small and
monotonic change affects only some certain elements. The effect is
sought from extremely the small expected change (of the order or less than 
$10^{-13} M_{\odot}$ per year), so it is sufficiently to consider the influence
within the framework of the two-body problem (the Sun and the planet), as it is 
usually done, because the correction due to the influence of other bodies in the weak 
effect associated with the change of the central mass, is still several orders of 
magnitude smaller. The two-body problem with the variable mass has a long standing
history and goes back to papers at the turn of the XIX and XX centuries by 
Gulden (1884), Meshchersky (1893), Stromgren (1903), Plummer (1906), etc. (see 
the detailed overview by Polyakhova, 1989). Variants of the isotropic mass 
variation in the two-body problem without the appearance of reactive forces, 
when no a particle is passed any pulse, were considered by MacMillan, Jeans, Armellini, 
Duboshin, Levi-Civita. A similar problem arises considering a possible change 
in time of the gravitational constant $ G (t) $ under the Dirac's hypothesis 
(Dirac, 1938) leads to the change of the mutual attraction forces and 
accelerations between bodies, and the analogous equations of 
the two-body problem. The analysis of equations for the central field of 
a variable mass body within the framework of General Relativity is given in the 
paper by Krasinsky and Brumberg (Krasinsky, Brumberg, 2004).

     If we denote $\mu(t)$ the product of $G(M_{\odot}+m)$, where
$m$ -- the planet mass, the vector equation of the relative motion for a body 
with mass $m$ is written
\begin{equation}\label{f-4}
 {\bf\ddot r} = - {\mu(t)\over{r^3}} \bf r .
\end{equation}
     In general, we can assume that the gravitational constant $G$, incoming 
in $\mu(t)$, may depend on time. Since the central field remains while  
$\mu(t)$ (\ref{f-4}) changes, then the area integral remains too, which 
is obtained immediately if the left and right side of (\ref{f-4}) to 
vector multiply on $\bf r$:
 $$ {\bf r}\times{\bf\ddot r}=0 \qquad \hbox{or} \qquad {d\over{dt}}({\bf r}\times\bf\dot r)=0,$$
then
\begin{equation}\label{f-5}
   {\bf r}\times\bf\dot r=\bf c.
\end{equation}
     The flat motion follows from the existence of the vector area integral 
(\ref{f-5}). Energy for the case of the time-dependent value of $\mu(t)$ is 
changing and is not more the integral of motion. Taking into account the 
monotony and smallness of the change of $\mu(t)$, it was shown (Jeans, 1924), 
that the invariant holds
\begin{equation}\label{f-6}
                 \mu(t)*a(t)=const ,
\end{equation}
where $a$ -- is the orbital semi-major planet axis.
    Sometimes this relation is called the adiabatic invariant of Poincare-Jeans, 
so as the first conclusion (\ref{f-6}) was made by Poincare (Poincar\'{e}, 
1911). The assumption, that $\mu(t)$ varies rather slowly, is essential for  
derivation of this result (Gelfgat, 1965). In our case, the expected value for 
Sun $\mid\dot \mu_{\odot}(t)/\mu_{\odot}(t)\mid \sim 10^{-13}$ per year is
much orders of magnitude smaller than it is required by the Gelfgat's
constraints.

     Since the plane of motion is preserved, then the elements ($i, \Omega$), 
determining the position of the orbital plane, do not change. It should be 
considered the dependence of semi-axis $a=a(t)$, the eccentricity $e=e(t)$ and 
the argument of pericentre $g=g(t)$ of the osculating elliptical orbits from the 
characteristics of the $\mu(t)$ change for instantaneous values of coordinates, 
velocities, and a mass. From (\ref{f-6}) we find a relationship 
between the changes of $\mu(t)$ and $a(t)$:
\begin{equation}\label{f-7}
 {\dot \mu \over \mu} = - {\dot a \over a}  .
\end{equation}

The relationship obtained from the integral of areas for the osculating orbit is
\begin{equation}\label{f-8}
       \mu(t)*a(t)*(1-e^2)= c^2 , \qquad \hbox{where} \qquad c=\mid \bf c\mid,
\end{equation}
therefore, using (\ref{f-6}), the $e$ eccentricity of the osculate orbit 
remains constant under the given conditions  $e=const$ (Jeans, 1925).
     Investigation of the change of the $g$ pericentre position for the small and
monotonic $\mu(t)$ change was made in the paper (Kevorkian, Cole, 1996), where it
was shown that under accepted conditions of the smallness and monotony of ${\dot \mu}$,
the $g$ value does not have a secular trend, and there may be only small
oscillations with the small amplitude of order $({\dot \mu}/\mu)^2$.

     The change in time of $a = a(t)$, as $\mu(t)$, leads to the change of the
period of the body $m$ that is leaving from the position for the case $\mu = const$, 
and the increase of deviation will depend quadratically on the time interval.

     The possible $\mu_{\odot}(t) = GM_{\odot}$ change in the Solar system should be
appeared in a systematic, progressive, although very small deviation of the body
position on the orbit (that is their longitude) and the change of the $a_i$ semi-major
axes proportionally to the $\mu_{\odot}(t)$ change with opposite sign (\ref{f-7}). 
The fact that the value $\mu(t) = G(M_{\odot} + m)$ includes the $m$ mass of a planet 
does not change the situation, since taking into account the mass $m$ of a 
planet leads to the correction by several orders less.

     Thus, when the area integral remains and the attractive force from the main body
decrease/increase monotone, the second body is moving along the trajectory
gradually receding/approaching from/to the central body. The relative
increase of the distance is equal to the relative decreasing of mass of the central
body and vice versa. An orbit is transforming gradually remaining identical to
itself,  and is of a spiral form. The $GM_{\odot}(t)$ change does not
lead to secular trends of the eccentricity and the longitude of perihelion. The
semi-axis is century changing $ a = a(t) $. Thus it is necessary to look for the effect
caused by possible change in time of the heliocentric gravitational constant
in the corresponding secular variation of the semi-axes of the planet orbits.

\bigskip
\centerline{\bf OBSERVATIONAL MATERIAL, REDUCTION OF OBSERVATIONS}
\smallskip

More than 635,000 positional observations of planets and spacecraft of various 
types (Table 1), mainly radiotechnical (1961-2010) have been used to construct 
high-precision ephemerides of planets and to determine the change of the 
heliocentric gravitational constant.
The very accurate observations are required to find the very small 
effects, and it is most important and desirable to have observational data for 
planets, close to the Sun and having shorter periods, in the first place, the 
data for Mercury and Venus. Radiotechnical measurements which began in 1961 and 
are continuing with rising numbers since, first, yielded two new types of 
measurements in astrometry: the distance and the relative speed, and secondly, 
the accuracy of the measurements became several orders of magnitude greater than 
the accuracy of the optical observations. 

\bigskip
{\bf Table 1.} The observations used

\bigskip
\begin{tabular}{c|c|c}
\noalign{\smallskip}
\hline
\noalign{\smallskip}
Observation type & Time interval & Observation number \\
\noalign{\smallskip}
\hline
\noalign{\smallskip}
Optical & 1913--2009 & 57768 \\
Radiotechnical & 1961--2010 & 577763  \\
\noalign{\smallskip}
\hline
\noalign{\smallskip}
Total &  & 635531 \\
\noalign{\smallskip}
\hline
\noalign{\smallskip}
\end{tabular}
\bigskip

\noindent For this reason, the ephemerides of 
the inner planets provided by radiotechnical observations (mostly, data of time 
delays) are based fully on these data. At present, the radiolocation of planet 
surfaces is not carried out, but trajectory data of various spacecraft that 
orbiting around planets or passing near them are received regularly. Accuracy 
of observations of ranging has improved from 6~km to several meters for today's 
data of the spacecraft. It is necessary to say that the ephemerides of the outer 
planets so far, mainly, are based on optical measurements since 1913, when at 
the Naval Observatory of USA the improved micrometer was introduced, and 
observations become more accurate ($0\rlap.''5$).

\bigskip
{\bf Table 2.} The distribution of optical observation and rms residuals in mas, 
1913--2009

\bigskip
\begin{tabular}{c|c|c}
\noalign{\smallskip}
\hline
\noalign{\smallskip}
Planet & Observation number & $\sigma$ \\ [4pt]
\noalign{\smallskip}
\hline
\noalign{\smallskip}
Jupiter &  13038 & 190 \\ [4pt]
Saturn &  16246 & 150 \\ [4pt]
Uranus & 11672 & 188 \\   [4pt]
Neptune & 11342 & 177 \\ [4pt]
Pluto & 5470 & 141 \\  [4pt]
\noalign{\smallskip}
\hline
\noalign{\smallskip}
Total &  57768 &   \\ [4pt]
\noalign{\smallskip}
\hline
\noalign{\smallskip}
\end{tabular}
\bigskip

\noindent However, until now a complete rotation period of Neptune and Pluto is 
not provided by the observations. In addition to optical observations of these 
planets, for the construction of ephemerides and estimation of their parameters 
the absolute observation satellites of the outer planets are used, as these 
observation are more precisely, and practically free from the phase effect hard 
taking into account, which is in observations of the planets themselves. Modern 
optical data are the CCD observations, their accuracy reaches $0\rlap.''05$. The 
number of used optical and radio observations, their planet distribution, as 
well as mean square error of residuals of observations are given in Table. 1 - 3.

\bigskip
{\bf Table 3.} Distribution, time interval, number and rms residuals for 
radiometric observations

\bigskip
\begin{tabular}{l|l|c|c|c}
\noalign{\smallskip}
\hline
\noalign{\smallskip}
Planet & Data type & Time interval & Observation number & $\sigma$ \\
\noalign{\smallskip}
\hline
\noalign{\smallskip}
Mercury & $\tau$ [m] & 1964--1997 & 937 &  575 \\[-2pt]
 & Š€ $\tau$ [m] &  1973--2009  & 5 &  31.0  \\[-2pt]
 & Š€ $\alpha,\delta$ [mas] & 2008--2009 & 6  & 2.1  \\[-2pt]
\noalign{\smallskip}
\hline
\noalign{\smallskip}
Venus  & $\tau$ [m]  & 1961--1995 & 2324 &  584 \\ [-2pt]
& Magellan  $dr$ [mm/s] &  1992--1994 & 195 &  0.007 \\ [-2pt]
& MGN,VEX  VLBI $\alpha,\delta$ [mas] & 1990--2007 & 22 &  3.0 \\[-2pt]
& VEX  $\tau$ [m] & 2006--2009 & 28163 &  3.6 \\[-2pt]
& Cassini  $\tau$ [m] & 1998--1999 & 2 &  2.4 \\[-2pt]
 & Cassini $\alpha,\delta$ [mas] & 1998--1999 & 4  & 105  \\[-2pt]
\noalign{\smallskip}
\hline
\noalign{\smallskip}
Mars   & $\tau$ [m]  & 1965--1995 & 54851 & 738 \\[-2pt]
&  Viking $\tau$ [m] &  1976--1982  & 1258 &  8.8  \\[-2pt]
&  Viking $d\tau$ [mm/s] &  1976--1978  & 14978 &  0.89  \\[-2pt]
&  Pathfinder $\tau$ [m] &  1997  & 90 &  2.8  \\[-2pt]
&  Pathfinder $d\tau$ [mm/s] &  1997  & 7569 &  0.09 \\[-2pt]
&  MGS $\tau$ [m] & 1998--2006 & 165562 &  1.4  \\[-2pt]
&  Odyssey $\tau$ [m] & 2002--2008 & 293707 &  1.2  \\[-2pt]
&  MRO $\tau$ [m] & 2006--2008 & 7775 &  1.6  \\[-2pt]
& spacecraft VLBI $\alpha,\delta$ [mas] & 1984--2010 & 136 &  0.6  \\[-2pt]
\noalign{\smallskip}
\hline
\noalign{\smallskip}
Jupiter & Š€ $\tau$ [m] &  1973--2000  & 7 &  13.8  \\[-2pt]
 & spacecraft $\alpha,\delta$ [mas] & 1973--2000 & 16  & 5.0  \\[-2pt]
& spacecraft VLBI $\alpha,\delta$ [mas] & 1996--1997 & 24 &  9.5  \\[-2pt]
\noalign{\smallskip}
\hline
\noalign{\smallskip}
Saturn & Š€ $\tau$ [m] &  1979--2006  & 34 &  3.5  \\[-2pt]
 & Š€ $\alpha,\delta$ [mas] & 1979--2006   & 92  & 0.4  \\[-2pt]
\noalign{\smallskip}
\hline
\noalign{\smallskip}
Uranus & Voyager-2 $\tau$ [m] &  1986  & 1  & 7.4  \\[-2pt]
 & Voyager-2 $\alpha,\delta$ [mas] &  1986  & 2  & 11.0  \\[-2pt]
\noalign{\smallskip}
\hline
\noalign{\smallskip}
Neptune & Voyager-2 $\tau$ [m] &  1989  & 1 &  22.9  \\[-2pt]
 & Voyager-2 $\alpha,\delta$ [mas] &  1989  & 2 &  3.7  \\[-2pt]
\noalign{\smallskip}
\hline
\noalign{\smallskip}
Total &  &  & 577763 \\
\noalign{\smallskip}
\hline
\noalign{\smallskip}
\end{tabular}
\bigskip

    The most accurate and long series of observations are available for Mars
for which spacecraft and landers were launched repeatedly. Radiotechnical 
observations relating to Venus are much smaller, there are spacecraft Magellan 
and Venus Express. A situation for observations of Mercury is much worse. The 
Messenger spacecraft (NASA) on the orbit around it have just appeared and we 
have not these data. There were only one-time conjunctions of Mariner-10 
(1974--1975) and Messenger (2008--2009) spacecraft and ranging for the Mercury 
surface (1964--1997). Situation should be changed after the new data from the 
Messenger spacecraft and from the future spacecraft BepiColombo (ESA, launch 2014)  
will be available. There are a number of radiotechnical 
observations for Jupiter and Saturn: the data for several spacecraft for 
Jupiter, and data of the Cassini spacecraft for Saturn. For Uranus and Neptune 
there are one 3-D points ($ \alpha, \delta, R $), resulting from the conjunction 
of Voyager-2 with these planets. The data were taken from the JPL database
(http:/ssd.jpl.nasa.gov/iau-comm4/), created by Dr. Standish, and now supported
by Dr. Folkner and the VEX data sent due to kindness of Dr. Fienga, as well as supplemented 
by rows of American and Russian radar observation of planets 1961--1995, taken 
from various sources. Russian radar observations of planets, along with their 
references are stored on the site of 
IAA RAS http://www.ipa.nw.ru/PAGE/DEPFUND/LEA/ENG/englea.htm. The brief 
description of astrometric radio observations can be found in Table.~2 
by Pitjeva (2005).

\bigskip
\centerline{\bf EPM2010 PLANETARY EPHEMERIDES, DETERMINED PARAMETERS}
\smallskip

     This work is based on the EPM2010 planetary ephemerides of IAA RAS. 
Numerical ephemerides of motion of the planets and the Moon (EPM --- Ephemerides 
of Planets and the Moon) began to create in the seventieth years of last century 
under the leadership of G.A. Krasinsky for support of Russian space flights, and 
has successfully developed since then. The version of the EPM2004 ephemerides 
has used to release the Russian ``Astronomical Yearbook'' described in (Pitjeva, 
2005), the version of the EPM2008 ephemerides in the paper (Pitjeva, 2010a). The 
EPM2010 ephemerides were constructed using more than 635 thousands of 
observations (1913-2010) of different types. EPM2010 differ from the previous 
versions by the improved dynamic model of motion of Solar system bodies, adding 
the perturbation from the ring of Trans-Neptunian objects (TNO), the new value 
of the Mercury mass, defined due to the three encounters of the Messenger 
spacecraft with Mercury, improvement of reductions of observations with the 
addition of the relativistic delay effect from Jupiter and Saturn, and the 
expanded database of observations, including radiotechnical (2008 - 2010) and 
CCD (2009) measurements.

     Ephemerides were constructed by the simultaneous numerical integration of 
equations of motion of all the major planets, the Sun, the Moon, the largest 301 
asteroids, 21 TNO, the lunar libration, taking into account the perturbations 
from the oblateness of the Sun and the asteroid belt, lying in the ecliptic 
plane and consisting of the remaining smaller asteroids, as well as the ring of 
the TNO rest at the mean distance of 43 au. The equations of motion of the 
bodies were taken in the post-Newtonian approximation in the Schwarzschild 
field. Integration in the barycentric system of coordinates for the epoch 
J2000.0 performed by the Everhard method over 400 years (1800-2200) by the lunar 
and planetary integrator of the software package ERA-7 (Krasinsky, Vasilyev, 
1997). The accuracy of numerical integration was verified by comparing the 
results of the forward and backward integrations over the century of the time 
interval. The errors were at least order of the magnitude smaller than the 
accuracy of observations. Thus, the accuracy of the ephemeris is determined 
mainly the accuracy of observations and their reductions.

     In the basic version of the improved EPM2010 planetary ephemeris about 260 
parameters are determined: elements of the orbits of the planets and the 18 
satellites of the outer planets; value of the astronomical unit; three angles of 
the orientation with respect to the ICRF frame; 13 parameters of the rotation of 
Mars and the coordinates for the three martian landers; masses of 10 asteroids, 
the mean density for the three taxonomic classes of asteroids (C, S, M), the 
mass and radius of the asteroid belt, the mass of the TNO ring, the ratio of the 
mass of the Earth and the Moon; quadrupole moment of the Sun ($ J_2 $) and 23 
parameters for the solar corona of different conjunctions of the planets with 
the Sun; eight coefficients of the topography of Mercury and the corrections to 
the level surface of Venus and Mars, constant shifts for the three series of 
planetary radar observations and for 7 spacecraft; 5 coefficients for the 
additional effect of the phase of the outer planets.

     The accuracy of EPM ephemerides was tested by comparison with the
observations (all residuals do not superior to their a priori errors), as well 
as comparison with the DE421 (JPL) independent ephemerides, those are in 
a good agreement (Pitjeva, 2010a).

     After constructing the EPM2010 ephemerides to all the observations the 
some other parameters can be to estimate: the changes of the $GM_{\odot}$ 
heliocentric gravitational constant, the $G$ gravitational constant, semi-axes 
of the planet orbits, and the astronomical unit.

\bigskip
\centerline{\bf OBTAINED ESTIMATES OF THE $\quad \bf M_{\odot}, G, a_i\quad$ 
CHANGES}
\smallskip

     The main problem of this case consists in the smallness of the effects
that need to be revealed. It was impossible to do this before an appearance in 
recent years a quite large number of high-precision observations, including data 
from spacecraft. Accuracy of determination of the parameters increased 
significantly also due to extension of the time interval for which there are 
high-precision sets of planet observations.

     The parameters $\dot G$ and $\dot {GM_{\odot}}$ were fitted by the
least squares method simultaneously with all basic parameters of ephemerides,
but each separately, i.e. they are considered in different solution versions.
If $\dot {GM_{\odot}}$ is found then it is taken into account
that the accelerations between the Sun and other bodies change, and the
mutual attractions between other pairs of bodies remain. This differs from the
situation when we find the change of $G$ and when the forces between all
bodies vary accordingly. It should be noted that for the version of the $\dot G$
definition from the planet motions, the main contribution makes by the Sun,
since the equations of the planet motions include products of the masses on
the gravitational constant, among them the term for the Sun ($GM_{\odot}$)
is the main one of several orders of magnitude more than the others.
Therefore, separating the change of $G$ from the change $M_{\odot}$ with the dominant term
of the $ GM_{\odot} $ is impossible. In this regard, it is more correctly (and reliably)
to determine from planet motions the change of $ GM_{\odot} $
instead of $ \dot G $ or $ \dot M_{\odot} $ separately.

\bigskip
{\bf Table 4.} The secular change values of the semi-major axes
for the 6 planets provided with the high-accuracy observations

\bigskip
\begin{tabular}{l|c|c}
\noalign{\smallskip}
\hline
\noalign{\smallskip}
Planet & $\dot a/a$ (century$^{-1}$)& Correlation coefficients \\ [4pt]
&& between $\dot a$ and $a$\\ [4pt]
\noalign{\smallskip}
\hline
\noalign{\smallskip}
  Mercury & (3.30 $\pm$ 5.95)$\cdot10^{-12}$& 56.5\%\\ [4pt]
  Venus & (3.74 $\pm$ 2.90)$\cdot10^{-12}$& 95.8\%\\ [4pt]
  Earth & (1.35 $\pm$ 0.32)$\cdot10^{-14}$& 0.6\%\\  [4pt]
  Mars  & (2.35 $\pm$ 0.54)$\cdot10^{-14}$& 0.4\% \\ [4pt]
  Jupiter & (3.63$\pm$ 2.24)$\cdot10^{-9}$& 20.2\%\\ [4pt]
  Saturn & (9.44$\pm$ 1.38)$\cdot10^{-10}$& 35.9\%\\ [4pt]
\noalign{\smallskip}
\hline
\noalign{\smallskip}
\end{tabular}
\bigskip

     Table. 4 shows the values obtained for the relative change of semi-major axes
of the planet orbits. The most accurate results connect with availability of
observations obtained with using radiotechnical equipment, in particular using
the spacecraft observations and the duration of larger time intervals. Accordingly,
the most reliable relative values of $ \dot a/a $ have been received for them.
These are the results for all the inner planets from Mercury to Mars. For
Jupiter and Saturn the accuracy of $ {\dot a/a} $ is less. Values for the planets,
not provided by radiotechnical data are unreliable. It is important that all the values
obtained for the planets from Mercury to Saturn show the positive values of the
$ \dot a/a $ ratio, i.e., indicate the decrease in time of the $ GM_{\odot}$ 
heliocentric gravitational constant  $(\ref{f-7})$.

  The change of the $ GM_{\odot} $ heliocentric gravitational constant has been
determined from fitting all observations:
\begin{equation}\label{f-9}
(\dot {GM_{\odot}}) /GM_{\odot} = (-5.04 \pm 4.14)\cdot10^{-14} \quad
\hbox{per year} \quad (3\sigma).
\end{equation}

This was made similarly to find $\dot G/G$:
\begin{equation}\label{f-10}
 \dot G/G = (-4.96 \pm 4.14)\cdot10^{-14} \quad \hbox{per year} \quad (3\sigma).
\end{equation}

     The closeness of the results (\ref {f-9}) and (\ref {f-10}) is not
surprising, since while finding the $ \dot G/G $, when the forces
between all pairs of bodies change, the effect of the central body is dominant
and again the effect of $ GM_{\odot} $ is found practically, instead of $G $.
   From the result (\ref {f-9}) obtained for $ GM_{\odot} $, it is possible to
estimate the $ \dot G $ value using the relation
\begin{equation}\label{f-11}
\dot \mu_{\odot} /\mu_{\odot}=\dot G/G+\dot M_{\odot}/M_{\odot}.
\end{equation}

This relation is valid with the 95$\%$ ($2\sigma$) probability:
\begin{equation}\label{f-12}
-7.8\cdot10^{-14}  < \dot G/G+\dot M_{\odot}/M_{\odot} < -2.3\cdot10^{-14}\quad \hbox{year}^{-1}.
\end{equation}

  Hence, using the limits (\ref {f-3}) found for the value
$ \dot M_{\odot}/M_{\odot} $, we obtain the $ \dot G/G$ value with the 95$\%$
probability is within the interval
\begin{equation}\label{f-13}
  -4.2\cdot10^{-14}  < \dot G/G < +7.5\cdot10^{-14}\quad \hbox{per year}.
\end{equation}

     Note, the $ \dot G/G $ estimate, obtained in 2004 from to the lunar laser 
ranging (Williams et al, 2004), which in any case is not complicated by the 
possible change of the solar mass, gives the following values of the limits for 
the gravitational constant change:
$ \dot G/G = (4 \pm 13) \cdot10^{-13} $ per year.
The estimate of $ (\dot {GM_{\odot}}) /GM_{\odot} $ obtained by us
(\ref {f-9}) has the opposite sign and its value is an order of
magnitude smaller.

The obtained change of $ GM_{\odot} $, most probably, is related to the change
of the Sun's mass $ M_{\odot} $, rather than to the $ G $ change. Thus, we have
\begin{equation}\label{f-14}
 \dot {M_{\odot}}/M_{\odot}  = (-5.0 \pm 4.1)\cdot10^{-14}\quad \hbox{per year} \quad (3\sigma).
\end{equation}
     Note that this value hits the limitation interval (\ref {f-3}) for
$ \dot {M_{\odot}}/M_{\odot}. $ The obtained change of $ GM_{\odot} $
(\ref {f-9}), possibly, reflects the balance between the mass loss due the radiation
and the solar wind and the falling material contained in comets, rocky debris
and asteroids, which do not produce the visible glowing gas tail.

\bigskip
\centerline{\bf THE POSSIBLE CHANGE OF THE ASTRONOMICAL UNIT}
\smallskip

   The change of the astronomical unit is connected with the change of
the heliocentric gravitational constant. The astronomical unit, although is
close in magnitude to the average distance of the Earth from the Sun,
but by its definition (resolution MAS 1976 - IAU, 1976) is connected with
the heliocentric gravitational constant:
\begin{equation}\label{f-15}
GM_{\odot} [m^3s^{-2}]=k^2\cdot \hbox{au}^3 [m^3]/86400^2[s^2],
\end{equation}
where k = 0.01720209895 is Gaussian gravitational constant. Currently, au
is determined from ranging data with very high real accuracy, allowing to deduce
the value of the heliocentric gravitational constant from the of value
$ {\hbox {au}~=} \ (149597870700 \, {\pm~3} $) m, using the relation
(\ref {f-15}): $ {GM_{\odot}~=} \, (1327124404 \, {\pm ~ 1}) [km^3s^{-2}] $
(e.g., Pitjeva and Standish, 2009), which coincides with the value $GM_{\odot}$, 
proposed by W.Folkner and obtained by the same method. These values of au and 
$ GM_{\odot} $ for the TDB time scale were approved at the XXVII IAU General 
Assembly (2009) as the best current values of astronomical constants
$(http://maia.usno.navy.mil/NSFA2/NSFA\_cbe.html)$.

     In the paper by Krasinsky and Brumberg (Krasinsky, Brumberg, 2004) from
raging data 1961 -- 2003, using a numerical theory of planetary motion, about 
coinciding with the EPM2004 (Pitjeva, 2005), the authors obtained the secular 
increase of the astronomical unit $ \dot {au} $ = 15 m per century, which
should correspond to the increase of the heliocentric gravitational constant
\begin{equation}\label{f-16}
\dot {GM_{\odot}} /GM_{\odot} \simeq 3 \cdot 10^{-12} \quad \hbox{per year}.
\end{equation}
     The positive change of au should correspond to the decrease of semi-major
axes of planet orbits, and not vice versa, as sometimes this is claimed,
and alternative theories of gravity (Miura T. et al,
2009; Nyambuya G.G., 2010) are even constructed on this incorrect basis.
Such the large positive change (\ref {f-16}) of
the  does not correspond to estimations of
physical processes in the Solar system (the solar radiation and wind, the matter
falling on the Sun), and also to the estimate (9) obtained in this study:
$ \dot {GM_{\odot}}/GM_{\odot} = -5.04 \cdot 10^{-14} $ per year.
However, authors  considered themselves that the increase of au and the
heliocentric gravitational constant (\ref {f-16}) are rather parameters of
agreement than the real change of the physical parameters.

     Analysis of the obtained results based on the observations described in
this paper, and the EPM2010 ephemerides, shows that the present level of observational
accuracy does not permit to evaluate the au change.
In the paper by Krasinsky and Brumberg the au change was
determined simultaneously with all other parameters, specifically, with the
orbital elements of planets and the value of the au astronomical unit itself.
However, at present it is impossible to determine simultaneously two parameters:
the value of the astronomical unit, and its change. In this case, the correlation
between $ au $ and its change $ \dot {au} $ reaches 98.1 $ \% $, and
leads to incorrect values of both of these parameters, in particular, gives
$ \dot {au} $ the order of 15 m per century.

     Without the simultaneous determination of $ au $ and
$ \dot {au} $, i.e. if only the change of the astronomical unit is estimated,
together with other parameters, the  $ \dot {au} $ value is about 1 m
per century, and does not exceed its formal uncertainty, thus it is determined:

$ \dot {au} $ = (1.2 $ \pm $ 3.2) m/cy (3 $ \sigma $).

     Furthermore, including or excluding the
$ \dot {au} $ value from the number of the solution parameters does not change
the observation residuals, the mean error of the unit weight is also not
changed ($ \Delta \sigma \simeq 0.2 \% $), so there is no reason to
assume that $ \dot {au} $ is the necessary parameter of agreement, and include it
in the number parameters to be estimated.

The modern accuracy has approached to the level when it is possible to estimate
the  change of the heliocentric gravitational constant $GM_{\odot}$, therefore
it is desirable to specify the definition of the astronomical unit, for example,
by fixing the connection between $GM_{\odot}$ and au at the certain time
moment, as it is inconvenient highly to have the changing value of the
astronomical unit.

\bigskip
\centerline{\bf CONCLUSION}
\smallskip

     Modern radiotechnical  observations of the planets and spacecraft having
the meter accuracy (relative error of $ 10^{-12} \div 10^{-11} $) make it possible
to obtain estimates of very small effects in the Solar system.
The significant progress is related to several factors: increase in
accuracy of reduction procedures for observations and in dynamical models of
planet motion, as well as improvement of the quality of observational data,
increasing their accuracy and the time interval in which these observations are
obtained.

     The results obtained indicate on the decrease of the heliocentric gravitation
constant per year at the level
$$ \dot {GM_{\odot}}/GM_{\odot}  = (-5.0 \pm 4.1)\cdot10^{-14} \quad (3\sigma).$$

     For the gravitation constant it is found that value of $ \dot G/G $ falls 
in the interval
 $$ -4.2\cdot10^{-14} < \dot G/G < +7.5\cdot10^{-14}$$
with the 95$ \% $ probability.

     The obtained change of $ GM_{\odot} $ seems to be due to the change of
the solar mass $ M_{\odot} $, rather than the $ G $ change and reflects
the balance between the loss of the solar mass due to by the radiation and
the solar wind and matter falling on the Sun.
It is possible to make the cautious conclusion that at present in the Solar system
there is still the significant effect of matter falling on the Sun,
that compensate partly the effect of reducing the solar mass due to the
radiation and the solar wind.

     In the future, the connection between $GM_{\odot}$ and au (15) should be
fixed at the certain time moment, as it is inconvenient highly
to have the changing value of the astronomical unit; moreover, it should be
to define the changes $ GM_{\odot},\, M_{\odot},\, G $ rather than the change 
of the astronomical unit.

\bigskip
\centerline{REFERENCES}
\smallskip
{
\parindent = 0.0truecm

\hangindent = 0.7truecm
\hangafter = 1
{\it Brandt, J.C.,} {\it Solnechnyi veter} (The Solar Wind), Moscow: Mir, 1973,
207~pp.

\hangindent = 0.7truecm
\hangafter = 1
{\it Gelfgat, B.E.,} K voprosu ob asimptoticheskom povedenii reshenii zadachi
dvuh tel peremennoi massy (On the asymptotic behavior of solutions of the
two-body problem with variable mass) // Trudy Astrofiz. in-ta AN KazSSR, 1965,
vol. 5, pp.~191--204.

\hangindent = 0.7truecm
\hangafter = 1
{\it Pitjeva E.V.,} High-precision ephemerides of planets -- EPM and determinations of some
astronomical constants // Solar System Research, 
2005, vol.~39, no. 3, pp.~176--186.

\hangindent = 0.7truecm
\hangafter = 1
{\it Polyahova E.N.,} Nebesnomehanicheskie aspekty zadach dvuh i treh tel s 
peremennymi massami (Sky-mechanical aspects of the problems of two- and 
three-body with variable mass) // Uchenye zapiski LGU, no. 424, Ser. matem.
nauk, Tr. Astr. obs., 1989, vol.~XLII, pp.~104--143.

\hangindent = 0.7truecm
\hangafter = 1
{\it Sunyaev R.A. et al.,} Fizika kosmosa (Space physics). Moscow: Sov. 
enciklopedia, 1986, 783~pp.

\hangindent = 0.7truecm
\hangafter = 1
{\it Hundhausen, A.J.,} {\it Rasshirenie korony i solnechnyi veter} (Coronal
Expansion and Solar Wind), Moscow: Mir, 1976, 302~pp.

\hangindent = 0.7truecm
\hangafter = 1
{\it Bailey, M.E., Chambers, J.E., and Hahn, G.,} Origin of sungrazers -
A frequent cometary end-state // Astron. Astrophys., 1992, vol.~257, no. 1,
pp.~315--322.

\hangindent = 0.7truecm
\hangafter = 1
{\it Brun, A.S., Turck-Chieze, S., Morel, P.,} Standard Solar Models in
the Light of New Helioseismic Constraints. I. The Solar Core // 
Astrophys. J., 1998, vol. 506, Issue 2, pp.~913--925.

\hangindent = 0.7truecm
\hangafter = 1
{\it Carroll, B.W., Ostlie, D.A.,} An introduction to modern astrophysics, 
Benjamin Cummings, 1996, 270~pp.

\hangindent = 0.7truecm
\hangafter = 1
{\it Dirac, P.A.M.,} A new basis for cosmology // Proc. Roy. Soc., London A, 
1938, vol.~165, no.~921, pp.~199--208.

\hangindent = 0.7truecm
\hangafter = 1
{\it Farinella, P., Froeschle, Ch., Gonzi, R., Hahn, G., Morbidelli, A.,
 Valsecchi, G.B.,} Asteroids falling onto the Sun // Nature, 1994, vol.~371,
pp.~314--317.

\hangindent = 0.7truecm
\hangafter = 1
{\it Frohlich, C., Lean, J.,} The sun's total irradiance: Cycles, trends and
related climate change uncertainties since 1976 // Geophys. Res. Lett., 1998,
vol.~25, pp.~4377--4380.

\hangindent = 0.7truecm
\hangafter = 1
{\it Frohlich, C., Lean, J.,} Solar radiative output and its variability:
evidence and mechanisms // Astron. Astrophys. Rev., 2004, vol.~12, pp.~273--320.

\hangindent = 0.7truecm
\hangafter = 1
{\it Gladman, B.J., Migliorini, F., Morbidelli, A. et al.,} Dynamical lifetimes 
of objects injected into asteroid belt resonances // Science, 1997, vol.~277,
pp.~197--201.

\hangindent = 0.7truecm
\hangafter = 1
{\it Hundhausen, A.J.,} Coronal mass ejections, in: Cosmic winds and the
heliosphere, Jokipii J.R., Sonett C.P., Giampapa M.S. (eds.). The University
of Arizona Press, Tucson, 1997, pp.~259--296.

\hangindent = 0.7truecm
\hangafter = 1
{\it Jeans, H.,} Cosmogonic problems associated with a secular decrease of mass //
 MNRAS, 1924, vol.~85, pp.~22.

\hangindent = 0.7truecm
\hangafter = 1
{\it Jeans, H.,} The effect of varying mass on binary system // MNRAS, 1925,
vol.~85, pp.~912--914.

\hangindent = 0.7truecm
\hangafter = 1
International Astronomical Union (IAU), ``Proceedings of the Sixteenth General Assembly,''
Transactions of the IAU, XVIB(31), pp.~52--66, (1976).

\hangindent = 0.7truecm
\hangafter = 1
{\it Kallenrode, M.-B.,} Space Physics: An Introduction to plasmas and particles
in the heliosphere and magnetospheres. Berlin: Springer, 2004, 482~pp.

\hangindent = 0.7truecm
\hangafter = 1
{\it Kevorkian, J.K., Cole, J.D.,} Multiple Scale and Singular Perturbation
Methods, Series: Applied Mathematical Sciences, New York: Springer, 1996, 648~pp.

\hangindent = 0.7truecm
\hangafter = 1
{\it Krasinsky, G.~A., Brumberg, V.A.,} Secular increase of astronomical unit from
analysis of the major planet motions, and its interpretation // Celest. Mech.
and Dyn. Astron., 2004, vol.~90, pp.~267--288.

\hangindent = 0.7truecm
\hangafter = 1
{\it Krasinsky, G.~A., Vasilyev, M.~V.,} ERA: knowledge
base for ephemeris and dynamical astronomy // Proc. of the IAU Coll.165,
Poznan, 1997, pp.~239--244.

\hangindent = 0.7truecm
\hangafter = 1
{\it Livingston, W.C.,}  Sun. In Allen's astrophysical quantities (Ed. A.N. Cox,
New York: Springer-Verlag, 2000), pp.~339--380.

\hangindent = 0.7truecm
\hangafter = 1
{\it Mann, I., Czechowski, A., Meyer-Vernet, N., et al.,} Dust in the
interplanetary medium // Plasma Phys. Control. Fusion., 2010, vol.~52 (124012), 10~pp.

\hangindent = 0.7truecm
\hangafter = 1
{\it Marsden, B.G.,} The sungrazing comet group. II. // Astron. J., 1989, vol.~98,
pp.~2306--2321.

\hangindent = 0.7truecm
\hangafter = 1
{\it Marsden, B.G.,} Sungrazing comets // Annual Rev. Astron. Astrophys., 2005, 
vol.~43, pp.~75--102.

\hangindent = 0.7truecm
\hangafter = 1
{\it Meyer-Vernet, N.,} Basics of the Solar Wind. Cambridge University Press,
2007, 463~pp.

\hangindent = 0.7truecm
\hangafter = 1
{\it Miura, T., Arakida, H., Kasai, M., Kuramata, S.,} Secular Increase of the
Astronomical Unit: a Possible Explanation in Terms of the Total Angular-Momentum
Conservation Law // Publications of the Astronomical Society of Japan, 2009, vol. 61,
no.~6, pp.~1247--1250.

\hangindent = 0.7truecm
\hangafter = 1
{\it Nyambuya, G.G.,} Azimuthally symmetric theory of gravitation - I. On
the perihelion precession of planetary orbits // MNRAS, 2010, vol.~403, pp.~1381--1391.

\hangindent = 0.7truecm
\hangafter = 1
{\it Pitjeva, E.V.,} EPM ephemerides and relativity // S. Klioner,
P.K. Seidelmann, M. Soffel (eds.), Proc. IAU Symp., no.~261, Relativity in
fundamental astronomy, Cambridge University Press, 2010a, pp.~170--178.

\hangindent = 0.7truecm
\hangafter = 1
{\it Pitjeva, E.V.,} Influence of asteroids and trans-neptunian objects
on the motion of major planets and masses of the asteroid main belt and the
TNO ring // Proc. Intern. conf. "Asteroid-comet hazard-2009", SPb.: Nauka,
2010b, pp.~237--241.

\hangindent = 0.7truecm
\hangafter = 1
{\it Pitjeva, E. V., Standish, E.M.,} Proposals for the masses of the three largest
 asteroids, the Moon-Earth mass ratio and the astronomical unit // Celest. Mech.
and Dyn. Astron., 2009, vol.~103, pp.~365-372.

\hangindent = 0.7truecm
\hangafter = 1
{\it Poincar\'{e}, H.,} Lecons sur les hypotheses cosmogoniques.
Paris: Gauthier-Villars, 1911, 294~pp.

\hangindent = 0.7truecm
\hangafter = 1
{\it Sekanina, Z., Chodas, P.W.,} Fragmentation hierarchy of bright sungrazing
comets and the birth and orbital evolution of the kreutz system. II. The Case
for Cascading Fragmentation // Astrophys. J., 2007, vol.~663. pp.~657--676.

\hangindent = 0.7truecm
\hangafter = 1
{\it Uzan, J.-Ph.,} The fundamental constants and their variation:
observational and theoretical status // Reviews of Modern Physics, 2003, vol.~75,
pp.~403--455.

\hangindent = 0.7truecm
\hangafter = 1
{\it Uzan, J.-Ph.,} Fundamental Constants and Tests of General
Relativity-Theoretical and Cosmological Considerations // Space Science Reviews,
2009, vol.~148, pp.~249--265.

\hangindent = 0.7truecm
\hangafter = 1
{\it Williams, J.G., Turyshev, S.G., Boggs, D.H.,}
Progress in Lunar Laser Ranging Tests of Relativistic Gravity.
// Physical Review Letters, 2004, vol.~93 (261101), 5~pp.

}
\end{document}